\documentstyle[aps,12pt]{revtex}

\title{ON THE PERTURBATIVE SOLUTIONS OF BOHMIAN QUANTUM GRAVITY}
\author{Fatimah Shojai\\Institute for Studies in Theoretical Physics and Mathematics,\\
P.O.Box 19395--5531, Tehran, IRAN.\\
Email: FATIMAH@NETWARE2.IPM.AC.IR}

\def\A{${\cal A}$}
\def\SS{${\cal S}$}
\def\Sc#1{${\cal S}_c^{(#1)}$}
\def\S#1{${\cal S}^{(#1)}$}
\def\OO{$\Omega$}
\def\O#1{$\Omega^{(#1)}$}
\def\I{$I$}
\def\II{$I\!\!I$}
\def\III{$I\!\!I\!\!I$}
\def\IIII{$I\!\!V$}
\def\MSS{{\cal S}}
\def\MSc#1{{\cal S}_c^{(#1)}}
\def\MS#1{{\cal S}^{(#1)}}
\def\MOO{\Omega}
\def\MO#1{\Omega^{(#1)}}
\def\MI{I}
\def\MII{I\!\!I}
\def\MIII{I\!\!I\!\!I}
\def\MIIII{I\!\!V}
\def\dsi{\frac{\delta\MS{0}}{\delta h_{ij}}}
\def\dsk{\frac{\delta\MS{0}}{\delta h_{kl}}}
\def\doi{\frac{\delta\MO{0}}{\delta h_{ij}}}
\def\dok{\frac{\delta\MO{0}}{\delta h_{kl}}}
\def\dssi{\frac{\delta\MS{2}}{\delta h_{ij}}}
\def\dssk{\frac{\delta\MS{2}}{\delta h_{kl}}}
\def\dooi{\frac{\delta\MO{2}}{\delta h_{ij}}}
\def\dook{\frac{\delta\MO{2}}{\delta h_{kl}}}
\def\dds{\frac{\delta^2\MS{0}}{\delta h_{ij}\delta h_{kl}}}
\def\ddo{\frac{\delta^2\MO{0}}{\delta h_{ij}\delta h_{kl}}}
\def\ddss{\frac{\delta^2\MS{2}}{\delta h_{ij}\delta h_{kl}}}
\def\ddoo{\frac{\delta^2\MO{2}}{\delta h_{ij}\delta h_{kl}}}
\def\dsp{\frac{\delta\MS{0}}{\delta \phi}}
\def\dssp{\frac{\delta\MS{2}}{\delta \phi}}
\def\dop{\frac{\delta\MO{0}}{\delta \phi}}
\def\doop{\frac{\delta\MO{2}}{\delta \phi}}
\def\ddsp{\frac{\delta^2\MS{0}}{\delta \phi^2}}
\def\ddssp{\frac{\delta^2\MS{2}}{\delta \phi^2}}
\def\ddop{\frac{\delta^2\MO{0}}{\delta \phi^2}}
\def\ddoop{\frac{\delta^2\MO{2}}{\delta \phi^2}}
\def\qehje{quantum Einstein--Hamilton--Jacobi equation}
\def\ehje{Einstein--Hamilton--Jacobi equation}
\def\be{\begin{equation}}
\def\ee{\end{equation}}

\begin{document}
\maketitle
\begin{abstract}
{\it In this paper we have solved the Bohmian equations of
quantum gravity, perturbatively. Solutions up to second order are derived
explicitly, but in principle the method can be used in any order. Some 
consequences of the solution are disscused.\\
{\bf PACS NO.:} 04.60.-m; 98.80.Hw; 03.65.BZ}
\end{abstract}
\section{INTRODUCTION}
Recently\cite{SAL,PAR,SOD}, a perturbative method for solving classical Einstein's equations
in its Hamilton--Jacobi form is presented. The method rests on expanding 
the Hamilton--Jacobi generating functional in terms of the powers of spatial 
gradiants of the metric and matter fields, and then solving the equations order
by order. This expansion is valid when the characteristic scale of spatial 
variation of physical quantities is larger than the characteristic lenght of
the theory, e.g. the Hubble's radius.
In fact it can be shown that the solution can be 
calculated at any order. The form of the Hamilton--Jacobi generating functional in each
order is chosen such that it be 3-diffeomorphic invariant. This method is
used and examined for many physical cases.
Salopek et. al.\cite{SAL} have solved the Hamilton--Jacobi and the momentum 
constraint equations in the presence of matter fields up to second order in 
spatial gradiants. Parry et. al.\cite{PAR} have used a specific conformal transformation
of 3--metric to simplify the Hamiltonian and solved the problem in higher orders
of spatial gradiants. Then they have compared their results with exact solutions 
for some specific cases and obtained a recursion relation for different orders
and so they have presented the solution up to any order. In addition similar calculations
are made for Brans--Dicke theory.\cite{SOD}

An essential question would be can the method be applied to quantum 
gravity realm. Unfortunately there are different approaches to quantum gravity,
non of them completely acceptable and self consistent. These include, the 
standard Wheeler--De Witt canonical approach\cite{DEW}, the Hawking path integral
approach\cite{HAR}, the Narlikar--Padmanabhan quantization of conformal degree of freedom
of the space--time metric\cite{NAR}, the Bohmian approach to quantum gravity\cite{HOL} and the approach
presented by author et.al. as geometrization of quantum theory\cite{GEO}. Among these
approaches Bohmian quantum gravity is of our concern here, becuase it highly
relates to Hamilton--Jacobi theory. In fact as we shall review in the next 
section, in Bohmian quantum gravity one encounters with a modified 
Hamilton--Jacobi equation.

We shall apply the above--mentioned perturbative method for solving Bohmian
quantum gravity equations. We shall do this up to the second order, but in
principle the metod can be applied to any order.
\section{BOHMIAN QUANTUM GRAVITY}
Bohm's theory is a causal version of quantum mechanics\cite{BOH}. According to this
theory, any particle is acompanied with an objectively real field ($\Psi$)
satisfying Schr\"odinger equation. This field exerts a quantum force derivable
from a quantum potential given by
\be
Q=-\frac{\hbar^2}{2m}\frac{\nabla^2|\Psi|}{|\Psi|}
\ee
This theory is motivated from the fact that when one sets 
$\Psi=|\Psi|\exp[i\MSS/\hbar]$ in the Schr\"odinger equation, one arrives at
a continuity equation:
\be
\frac{\partial |\Psi|^2}{\partial t}+\vec{\nabla}\cdot\left ( |\Psi|^2
\frac{\vec{\nabla}\MSS}{m}\right )=0
\ee
and a modified Hamilton--Jacobi equation:
\be
\frac{\partial \MSS}{\partial t}+\frac{|\vec{\nabla}\MSS|^2}{2m}+V+Q=0
\label{oh}
\ee
in which $V$ represents the classical potential and $Q$, the quantum potential 
is defined as above. It is the main positive point of Bohm's theory which
using only quantum potential is able to explain all enigmatic aspects of quantum
theory. These includes presentation of a causal description for wave--function
collapse during a measurement, and description of uncertainty relations and also
presentation of particle trajectories\cite{BOH}. Particle trajectory can be obtained through the modified Hamilton--Jacobi equation (\ref{oh}) and using the guidance relation $\vec{p}=\vec{\nabla}{\cal S}$, or using the Newton's law of motion including the quantum potential. It is worth noting that the trajectories explain many nonordinary behaviour in quantum mechanics. For example particle trajectories in a two--slit experiment can be calculated\cite{HOL} and it can be seen how quantum potential forces particles to move in such a way to make the interference pattern.

Bohm's theory can be applied to any system. Application of this theory to
gravity leads to Bohmian quantum gravity. Its properties and positive points 
are expressed in the literature\cite{HOL,BOH}.
Application of Bohm's theory to quantum gravity has several advantages. First
of all in this approach, different quantities like the 3--space geometry, 
intrinsic and extrinsic curvatures of the space--like surfaces and so on have
physical reality without any dependence upon the measurment process. Second,
the metric has a definite time evolution in this theory. 
Third, in this approach the wave function has two roles. One role in generating 
the quantum potential and another as the probabilistic interpretation. When
one deals with a single system (as is the case for quantum cosmology)
 for which the probability is not defined, the 
first role of the wave function is important. Note that in the standard quantum
theory, only the second role is highlighted and thus the meaning of the wave function
is questionable in quantum gravity.
Finally, the classical
limit is well defined in Bohm's theory. When both quantum potential and its gradiant
are small compared to the classical potential and its gradiant, then we are in the
classical limit. This allows one, in specific cases, to have quantum effects at large scales and 
classical limit in small scales. 

Here we use Bohmian quantum gravity, not only because of the above mentioned
advantages, but also because it highly relates to the Hamilton--Jacobi equation.
Before proceeding, we present the Bohmain equations for quantum gravity\cite{HOL}
which we shall refer to later. These equations are\footnote{They can be obtained by setting $\Psi={\cal A}\exp[i{\cal S}/\hbar]$ in the WDW equation and the 3--diffeomorphism invariance condition.}  (setting $\hbar=c=8\pi G=1$):
\be 
\frac{\delta}{\delta h_{ij}}\left ( 2h^q G_{ijkl}\frac{\delta {\cal S}}{\delta h_{kl}}
{\cal A}^2\right )+\frac{\delta}{\delta\phi}\left (\frac{h^q}{\sqrt{h}}\frac{\delta {\cal S}}{\delta \phi} {\cal A}^2
\right )=0
\label{CON}
\ee
\be
G_{ijkl}\frac{\delta {\cal S}}{\delta h_{ij}}\frac{\delta{\cal S}}{\delta h_{kl}}
+\frac{1}{2\sqrt{h}}\left (\frac{\delta{\cal S}}{\delta\phi}\right )^2
-\sqrt{h}\left ({\cal R}^{(3)}+2\Lambda-Q_G\right )+\frac{1}{2}\sqrt{h}
h^{ij}\partial_i\phi\partial_j\phi+\frac{1}{2}\sqrt{h}(V+Q_M)=0
\label{HAM}
\ee
\be
Q_G=-\frac{1}{\sqrt{h}{\cal A}}\left ( G_{ijkl}\frac{\delta^2{\cal A}}{\delta h_{ij}\delta h_{kl}}
+h^{-q}\frac{\delta h^qG_{ijkl}}{\delta h_{ij}}\frac{\delta {\cal A}}{\delta h_{kl}}
\right )
\ee
\be
Q_M=-\frac{1}{h{\cal A}}\frac{\delta^2{\cal A}}{\delta \phi^2}
\ee
\be
2\nabla_j\frac{\delta {\cal A}}{\delta h_{ij}}-h^{ij}\partial_j\phi\frac{\delta {\cal A}}{\delta \phi}=0
\label{X}
\ee
\be
2\nabla_j\frac{\delta {\cal S}}{\delta h_{ij}}-h^{ij}\partial_j\phi\frac{\delta {\cal S}}{\delta \phi}=0
\label{Y}
\ee
in which ${\cal A}$ is the norm of the wave function, ${\cal S}$ is its phase times $\hbar$
and is in fact the quantum Einstein--Hamilton--Jacobi function, $q$ is an
ordering parameter, $h_{ij}$ is the spatial metric in ADM decomposition
of the space--time metric, $G_{ijkl}$ is super metric on 3--space,
 $\phi$ denotes the
matter field, and $Q_G$ and $Q_M$ are gravity and matter quantum potentials
respectively.

Equation (\ref{CON}) is the continuity equation representing the conservation
law of probability in the super space, and equation (\ref{HAM}) is 
the quantum Einstein--Hamilton--Jacobi equation, which shows that the difference
between quantum and classical worlds is only the presence of the quantum potential
consisting of two terms, gravity and matter quantum potentials.
Equations (\ref{X}) and
(\ref{Y}) are 3-diffeomorphism invariance conditions for ${\cal A}$ and ${\cal S}$.
Time evolution of metric and the matter field can be derived from the canonical
relations:
\be
\pi^{kl}=\frac{\delta{\cal S}}{\delta h_{kl}}=\frac{\sqrt{h}}{2}(K^{kl}-h^{kl}K)
\ee
\be
\pi_\phi=\frac{\delta{\cal S}}{\delta\phi}=\frac{\sqrt{h}}{N}\dot{\phi}
-\sqrt{h}\frac{N^i}{N}\partial_i\phi
\ee
\be
K_{ij}=\frac{1}{2N}(\nabla_iN_j+\nabla_jN_i-\dot{h}_{ij})
\ee
in which $N$ and $N^i$ are the lapse and shift functions respectively, and 
$K_{ij}$ is the extrinsic curvature of the 3--space.  It can be seen that in Bohmian quantum gravity, there is no time problem. Time emerges naturally from the equations of motion. Bohmian trajectories can be obtained from the above equations. For example, the Bohmian trajectories for Rabertson--Walker universe are derived by Horiguchi in \cite{HOL} and other references cited in \cite{HOL}. Another example is Bohmian trajectories for black holes. They are obtained in \cite{KEN}. In this reference it is shown that the quantum black hole geometry is highly sensible to the ordering parameter. For some specific ordering parameter, Bohmian quantum gravity presents a good framework for understanding Hawking radiation.  Some other aspects of Bohmian quantum gravity can be found in \cite{VIN}. For a complete review of the theory see \cite{HOL}.
\section{SOLVING THE EQUATIONS}
It is disscussed in the previous section that the complete set of equations
of quantum gravity are equations (\ref{CON}), (\ref{HAM}), (\ref{X}), 
and (\ref{Y}). The first is the continuity equation,
while the second is the \qehje .
 The third and fourth equations gurantee that \A\ and \SS\ be 3-diffeomorphic
invariants. A perturbative solution can be achieved via expansion of \SS\ and
\A\ in terms of powers of spatial gradiants. In the long--wavelength approaximation
a few terms of the expansion is sufficient. Therefore one should set:
\begin{eqnarray}
&&\MOO=\sum_{n=0}^\infty \MO{2n} ; \ \ \ \ {\cal A}=e^\MOO \\
&&\MSS=\sum_{n=0}^\infty \MS{2n}
\end{eqnarray}
Note that introducing the new functional \OO\ will simplifies the equations.
In each order, the two coupled equations \qehje\ and continuity equation
should be solved. The two other equations only show that the functionals
\S{2n}\ and \O{2n}\ must be 3-diffeomorphic invariants. By considering
special forms for \S{2n}\ and \O{2n}, these equations would be satisfied
automatically. 
\subsection{Zeroth Order Solution}
In this order, the continuity equation reads as:
\[ -\left ( q+\frac{3}{2}\right )h_{ij}\dsi+4\sqrt{h}G_{ijkl}\doi\dsk \]
\be
+2\sqrt{h}G_{ijkl}\dds+2\dop\dsp+\ddsp=0
\label{C0}
\ee
while the zeroth order \qehje\ is:
\[ 2\sqrt{h}G_{ijkl}\dsi\dsk+\left (\dsp\right )^2-2\sqrt{h}G_{ijkl}\ddo \]
\be
-2\sqrt{h}G_{ijkl}\doi\dok+\left ( q+\frac{3}{2}\right )h_{ij}\doi
-\ddop-\left (\dop\right )^2=0
\label{H0}
\ee
in which for simplicity of calculations, we have assumed that the scalar field
has no self interaction, i.e we have set $V(\phi)=0$.

In order to \S{0}\ and \O{0}\ be 3-diffeomorphic invariants and thus satisfy
equations (\ref{X}) and (\ref{Y}) automatically, one should set:
\be
\MS{0}=\int\!d^3x \sqrt{h}H(\phi)
\ee
\be
\MO{0}=\int\!d^3x \sqrt{h}K(\phi)
\ee
in which $H$ and $K$ are functions of the scalar field and contain no spatial
derivatives. Since $d^3x\sqrt{h}$ is 3-diffeomorphic invariant measure, the
above experssions are also 3-diffeomorphic invariant. By substituting these relations
for \O{0} and \S{0} in equations (\ref{C0}) and (\ref{H0}), we have the following
equations for $H$ and $K$:
\be
\frac{d^2H}{d\phi^2}-\frac{3}{2}(q+5)H+2\sqrt{h}\left ( \frac{dH}{d\phi}
\frac{dK}{d\phi}-\frac{3}{4}KH\right )=0
\ee
\be
\frac{d^2K}{d\phi^2}-\frac{3}{2}(q+5)K-\sqrt{h}\left ( \left (\frac{dH}{d\phi}\right )^2
-\left (\frac{dK}{d\phi}\right )^2+\frac{3}{4}K^2-\frac{3}{4}H^2\right )=0
\ee
Setting both metric--dependent (terms containing $\sqrt{h}$) and 
metric--independent terms equal to zero, we have four equations with the
simultaneous solution:
\begin{eqnarray}
H&=&Ae^{\alpha\phi};\ \ \ \ \alpha=\pm\frac{\sqrt{3}}{2} \\
K&=&BH\\
q&=&-\frac{9}{2}
\end{eqnarray}
in wich $A$ and $B$ are constants of integration. It must be noted here that
using this solution it is a simple task to show that quantum potential is
zero at this order. So the solution at this order is in fact classical.
\subsection{Second Order Solution}
In the second order, the continuity and \qehje s are respectively:
\[ -\left ( q+\frac{3}{2}\right )h_{ij}\dssi+4\sqrt{h}G_{ijkl}\doi\dssk
+4\sqrt{h}G_{ijkl}\dooi\dsk\]
\be
+2\sqrt{h}G_{ijkl}\ddss+2\dop\dssp+2\doop\dsp+\ddssp=0
\ee
\[ 2\sqrt{h}G_{ijkl}\dsi\dssk+\dsp\dssp-\sqrt{h}G_{ijkl}\ddoo-2\sqrt{h}G_{ijkl}
\doi\dook\]
\be
+\frac{1}{2}(q+3)h_{ij}\dooi-\frac{1}{2}\ddoop-\dop\doop-\sqrt{h}\left (
{\cal R}^{(3)}-\frac{1}{2}\nabla_i\phi\nabla^i\phi\right )=0
\ee
On using the zeroth order solution and again setting both terms with and without
$\sqrt{h}$ equal to zero, one arrives at the following four equations:
\be
\MI[\MS{2}]\equiv2\sqrt{h}G_{ijkl}\ddss+3h_{ij}\dssi+\ddssp=0
\ee
\be
\MII[\MS{2},\MO{2}]\equiv 2\alpha B\dssp-Bh_{ij}\dssi+2\alpha\doop-h_{ij}\doop=0
\ee
\[ \MIII[\MS{2},\MO{2}]\equiv Hh_{ij}\dssi-2\alpha H\dssp \]
\be
-BHh_{ij}\dooi+2\alpha HB\doop
+2\sqrt{h}\left ( {\cal R}^{(3)}-\frac{1}{2}\nabla_i\phi\nabla^i\phi\right )=0
\ee
\be
\MIIII[\MO{2}]\equiv 2\sqrt{h}G_{ijkl}\ddoo+3h_{ij}\dooi+\ddoop=0
\ee
In order to solve the above equations, we use a different method with respect to 
the zeroth order. Our goal is to find the quantum corrections on the Hamilton--Jacobi
functional at second order, to the classical functional \Sc{2}.
It must be noted that \III$[$\S{2},\O{2}$]=0$ is just the classical
\ehje\ except for its third and fourth terms. So, with a glance at the form of the
third and fourth terms, one easily can solve \III$=0$ as:
\be
\MS{2}-B\MO{2}=\MSc{2}
\label{Z}
\ee
Therefore for finding \S{2}, it is sufficient to solve \IIII$[$\O{2}$]=0$
to find \O{2}\ and use the above equation. On the other hand, since for the classical
limit the $\sqrt{h}$--independent terms of the continuity equation leads to 
\I$[$\Sc{2}$]=0$ and since \I\ is linear, we have:
\be
\MI[\MS{2}]=\MI[\MSc{2}+B\MO{2}]=\MI[\MSc{2}]+B\MI[\MO{2}]=B\MI[\MO{2}]=B\MIIII[\MO{2}]
\ee
so
\be 
\MI[\MS{2}]=0 \Longleftrightarrow \MIIII[\MO{2}]=0
\ee
It remains for the second equation \II$[$\S{2},\O{2}$]=0$. On using the 
relation (\ref{Z}), and linearity of \II, one arrives at:
\be
2\alpha\doop-h_{ij}\dooi=\frac{B}{1+B^2}\left ( -2\alpha \frac{\delta\MSc{2}}{\delta\phi}
+h_{ij}\frac{\delta\MSc{2}}{\delta h_{ij}}\right )
\ee
which has the solution:
\be
\MO{2}=-\frac{B}{1+B^2}\MSc{2}+\Lambda
\label{P}
\ee
where the functional $\Lambda$ satisfies the equation:
\be
2\alpha\frac{\delta\Lambda}{\delta\phi}=h_{ij}\frac{\delta\Lambda}{\delta h_{ij}}
\label{L}
\ee
In addition, using the relation (\ref{P}) and \IIII$[$\O{2}$]=0$ and the linearity
of \IIII\ one has \IIII$[\Lambda]=0$. So it is sufficient to find the simultaneous 
solution of the relations (\ref{L}) and \IIII$[\Lambda]=0$. In finding the solution,
we use the techniques of \cite{PAR}. Making the conformal transformation:
\be
f_{ij}(x)=F^{-2}[\phi(x)]h_{ij}(x)
\ee
one can see that the equation (\ref{L}) requires $F$ to satisfy the relation:
\be
-4\alpha\frac{dF}{d\phi}=F
\ee
with the solution $F=constant\times \exp[-\phi/4\alpha]$. The most general
form of $\Lambda$ is 
\be
\Lambda=\int\!d^3x\sqrt{f}\left [ L(\phi)\tilde{\cal R}^{(3)}+M(\phi)\tilde{\nabla}_i\phi
\tilde{\nabla}^i\phi\right ]
\ee
where a tilde over any quantity represents that it is calculated using the
$f_{ij}$ metric. $\tilde{\cal R}^{(3)}$ is the Ricci scalar curvature of
$f_{ij}$, $L$ and $M$ are some functions of the scalar field. The above
expression is the most general form to make $\Lambda$, 3--diffeomorphic invariant
and contains terms with spatial gradiants of order two. Note that terms
like $\tilde{\nabla}^2\phi$ can be transformed to $\tilde{\nabla}_i\phi
\tilde{\nabla}^i\phi$ by integration by part. 

Now the equations (\ref{L}) and \IIII$[\Lambda]=0$ can be solved for $L$ and
$M$. The solution can be transformed back to the original metric $h_{ij}$
using the inverse of the above conformal transformation. The result is:
\be
\Lambda=C\int\!d^3x\sqrt{h}e^{\phi/4\alpha}\left [{\cal R}^{(3)}-\frac{1}{6}
\nabla_i\phi\nabla^i\phi\right ]
\ee
where $C$ is a constant.

For writting down \S{2}\ and \O{2}\ it is neccesary to know \Sc{2}. From
\cite{PAR}, we have:
\be
\MSc{2}=\frac{3}{10}\int\!d^3x\sqrt{h}e^{\alpha\phi}
\left [{\cal R}^{(3)}-\nabla_i\phi\nabla^i\phi\right ]
\ee
Thus we have:
\[ \MSS=\MS{0}+\MS{2}+\cdots=\int\!d^3x\sqrt{h}\left \{e^{\alpha\phi}
\left [ A+\frac{3}{10(1+B^2)}\left ({\cal R}^{(3)}-\nabla_i\phi\nabla^i\phi
\right )\right ] \right.\]
\be
\left . +BC e^{\phi/4\alpha}\left ({\cal R}^{(3)}-\frac{1}{6}
\nabla_i\phi\nabla^i\phi\right )\right \}\cdots
\ee
\[ \MOO=\MO{0}+\MO{2}+\cdots=\int\!d^3x\sqrt{h}\left \{e^{\alpha\phi}
\left [ AB-\frac{3B}{10(1+B^2)}\left ({\cal R}^{(3)}-\nabla_i\phi\nabla^i\phi
\right )\right ] \right.\]
\be
\left . +C e^{\phi/4\alpha}\left ({\cal R}^{(3)}-\frac{1}{6}
\nabla_i\phi\nabla^i\phi\right )\right \}\cdots
\ee
It is worth noting that the first two terms in ${\cal S}$ which are scaled by $e^{\alpha\phi}$ are of the same form as the classical solution up to second order. In fact quantum effects are introduced via the third term and the renormalization of the factor $3/10$ in the second term to $3/10(1+B^2)$. An important property of the solution is the factor $e^{4\phi/\alpha}$ of the third term which differs from the factor of the first two terms. The presence of the third term leads to new couplings between the matter field and the metric in Bohmian equations of motion, leading to highly quantic solutions.
\section{CONCLUDING REMARKS}
As we saw, the Bohmian equations of motion for quantum gravity, i.e.
\qehje\ and continuity equation can be solved in principle as an 
expansion with respect to spatial gradiants. We derived the solution 
up to the second order. As a result since our solution contains spatial
gradiants, it is useful for disscusing inhomogeneous space--times such
as black holes which are partly disscused in the framework of 
Bohmian quantum gravity\cite{KEN}. In a forthcomming paper we shall apply the 
result to black holes.

A point must be noted here. As we asserted previousely, according to
Bohm's theory, in
the classical limit quantum potential and its gradiant are small compared
to classical potential and its gradiant. This can be achieved both in the case
where the norm of the wave function varies slowly and in the case where it
varied highly. This is because of the fact that quantum potential is proportional
to the fraction of second derivaties of the norm of the wave function and the norm itself.
(see e.g. \cite{HOL}). As a result, classical limit and long wavelenght limit
(i.e. considering only a few terms in the expansion with respect to spatial gradiants)
are not the same. So in Bohmian quantum gravity comparison of characteristic lenght
of fluctuations with theory's characteristic lenght (e.g. Hubble's radius)
does not lead us to anything about the fact that the limit is either classic or
quantum.

\end{document}